\documentstyle[prl,aps,epsfig,floats]{revtex}

\begin{document}


\wideabs{

\title{\bf  Experimental Realization of a 2D Fractional Quantum Spin Liquid }

\author{R. Coldea$^{1,2}$, D.A. Tennant$^{2,3}$, A.M. Tsvelik$^4$,
and Z. Tylczynski$^5$}

\address{
$^1$Oak Ridge National Laboratory,
Oak Ridge, Tennessee 37831, USA\\
$^2$ISIS Facility, Rutherford Appleton Laboratory,
Chilton, Didcot, Oxon OX11 0QX, UK\\
$^3$Oxford Physics, Clarendon Laboratory, Parks Road,
Oxford OX1 3PU, UK\\
$^4$University of Oxford, Department of Theoretical Physics,
1 Keble Road, Oxford OX1 3NP, UK\\
$^5$Institute of Physics, Adam Mickiewicz University, 
Umultowska 85, 61-614 Poznan, Poland}

\date{3 November 2000}

\maketitle

\begin{abstract}
The ground-state ordering and dynamics of the two-dimensional
(2D) $S$=1/2 frustrated Heisenberg antiferromagnet Cs$_2$CuCl$_4$
is explored using neutron scattering in high magnetic fields. We
find that the dynamic correlations show a highly dispersive 
continuum of excited states, characteristic of the RVB state,
arising from pairs of $S$=1/2 spinons. Quantum renormalization
factors for the excitation energies (1.65) and incommensuration
(0.56) are large.
\end{abstract}

 } 

\narrowtext

The concept of fractional quantum states is central to the modern theory of
strongly correlated systems. In magnetism, the most famous example is the
spin $S$=1/2\ 1D Heisenberg antiferromagnetic chain (HAFC) where pairs of 
$S$=1/2\ spinons are deconfined from locally allowed $S$=1 states; a phenomenon
that is now well established both theoretically \cite{Muller81} and
experimentally \cite{Tennant95a}. These spinons are topological excitations
identified with quantum domain walls. Experimentally, such fractionalization
is manifest as a highly dispersive continuum in the dynamical magnetic
susceptibility measured by {\it e.g.} neutron scattering \cite{Tennant95a},
and for the HAFC identified as creation of pairs of spinons.

In 1973 Anderson \cite{Anderson73} suggested that a 2D fractional quantum
spin liquid may take the form of a ``resonating valence bond'' (RVB) state 
comprising singlet spin pairings in the ground state, and with pairs of
excited $S$=1/2 spinons separating via rearrangement of those bonds. The
dominant feature of the RVB state, present in all its theoretical
descriptions \cite{Kalmeyer87,larkin,lee} is an extended, highly-dispersive,
continuum. To date this feature remains unobserved in any 2D magnet; in the
case of the $S$=1/2 Heisenberg square lattice (HSL) mean field confining
effects lead to $S$=1 magnons and a renormalized classical picture of
fluctuations around local N\'eel order emerges \cite{Hayden91,Sachdev99}.
One may think, however, that because frustrating interactions can counteract
the staggered fields responsible for confinement \cite{Sachdev99,Watanabe99}, 
they may provide a route to generating fractional phases in 2D.

We explore such a scenario by making neutron scattering studies on 
Cs$_{2}$CuCl$_{4}$. By exploiting its unique experimental properties as a
low-exchange quantum magnet \cite{Coldea97} we reveal an unexpectedly strong
two-dimensionality in the form of a triangular antiferromagnet with
partially released frustration. The simplicity of the couplings in 
Cs$_{2}$CuCl$_{4}$ makes it a model system to investigate generic features 
of 2D frustrated quantum antiferromagnets.

The structure of Cs$_{2}$CuCl$_{4}$ is orthorhombic ({\em Pnma}) with
lattice parameters $a$=9.65 \AA , $b$=7.48 \AA\ and $c$=12.35 \AA\ at 0.3 K.
Magnetic interactions are mostly restricted between Cu$^{2+}$\ $S$=1/2
spin-sites in the ($b,c$) plane, see Fig.\ \ref{fig1}(a), with coupling $J$
along $b$ (``chains'') and zig-zag ``interchain'' coupling $J^{\prime }$
along the $c$-axis \cite{Coldea96}. A small interlayer coupling $J^{\prime
\prime }<10^{-2}J$\ (along $a$) stabilizes 3D order below $T_{N}=0.62$ K
into an incommensurate structure along $b$ due to the frustrated couplings;
weak anisotropies confine the ordered moments to rotate in cycloids near
coincident with the ($b,c$) plane, see Fig.\ \ref{fig1}(d) but with a small
tilt of the cycloidal plane relative to the ($b,c$) plane whose sense
alternates along $c$ such that for each plaquette (isosceles triangle) $%
\langle {\bf S}_{1}\cdot ({\bf S}_{2}\times {\bf S}_{3})\rangle $ is small
but nonzero (order is noncoplanar) \cite{Coldea96}, making this system a 
candidate for a chiral spin state \cite{Wen89}. The minimal Hamiltonian 
determining the magnetic order is 
\begin{equation}
{\cal H}=\sum\limits_{\langle i,i^{\prime }\rangle }J{\bf S}_{i}\cdot {\bf S}%
_{i^{\prime }}+J^{\prime }\sum\limits_{\langle i,j\rangle }{\bf S}_{i}\cdot 
{\bf S}_{j}  \label{hamiltonian}
\end{equation}
with each interacting spin-pair counted once, see Fig.\ \ref{fig1}(a); a
detailed description of the full Hamiltonian including small
Dzyaloshinskii-Moriya terms is given elsewhere \cite{longpaper}. The
Hamiltonian interpolates between non-interacting HAFCs ($J^{\prime }=0$),
the fully frustrated triangular lattice ($J^{\prime }=J$), and unfrustrated
HSL ($J=0$).

Quantifying the couplings in (1) is of considerable importance both to guide
theory and put our results in context. We do this using the following
approach:\ neutron diffraction measurements were made on a single crystal of
Cs$_{2}$CuCl$_{4}$ in magnetic fields up to 7 T and temperatures down to 0.2
K using the PRISMA time-of-flight (TOF) spectrometer at the ISIS spallation
neutron source. For fields along $a$ (near perpendicular to the planes of 
spin rotation) a 3D incommensurate ``cone'' order is stable up to full
ferromagnetic (F) alignment ($B_{c}=8.44$ T at $T$=0.03 K), see Fig.\ \ref
{fig1}(c). At $T$=0.2 K magnetic Bragg peaks arising from the 
{\em transverse} spin rotation move from $\epsilon _{0}$=0.030(2) in 
zero field to $\epsilon $=0.047(2) at 7 T, where $\epsilon$ is the 
incommensuration relative to N\'{e}el order, see Fig.\ \ref{fig1}(f).
Mean-field theory predicts no change with field, and the large
renormalization observed is a purely quantum effect. Since the ferromagnetic
(F) state is an eigenstate of (\ref{hamiltonian}) with no fluctuations, 
$\epsilon$ at the saturation field $B_{c}$ is at its classical value 
\cite{Weihong99} $\sin {\pi \epsilon _{c}}=J^{\prime }/2J$. 
Higher-field measurements \cite{longpaper} observe $\epsilon_{c}=0.053(1)$, 
implying an exchange coupling ratio of $J^{\prime }/J$=0.33(1). The 
resulting quantum renormalization of the zero-field incommensuration 
$\epsilon _{0}/\epsilon _{c}$=0.56(2) is similar to the predicted value 
of 0.43(1) ($J^{\prime }/J$=0.33(1)) estimated by series expansions 
using a paired singlet basis \cite{Weihong99}
Additionally, the determined exchange coupling ratio is in agreement with 
the observed 2D dispersion in the saturated phase at 12 T $\parallel a$ 
\cite{longpaper}, which give the bare exchange couplings-per-site 
$J$=0.375(5) meV within chains and $2J^{\prime}$=0.25(1) meV between chains. 
{\em This demonstrates that ``interchain'' couplings are of the same order 
as ``intrachain'' and Cs$_{2}$CuCl$_{4}$ is therefore a quasi-2D system}.
These observations require a change in the point of view taken by earlier 
studies \cite{Coldea97}, which proposed a quasi-1D picture based on 
estimates of the ``interchain'' couplings not including the large quantum 
renormalization of the incommensuration reported here. We now present 
detailed measurements of the excited states.

Dynamical correlations in a 2.5 cm$^{3}$ single crystal of Cs$_{2}$CuCl$_{4}$
were probed using the indirect-geometry TOF spectrometer IRIS, also at ISIS.
The energy resolution ($0.016$ meV full-width-half-maximum (FWHM)) was an
improvement of nearly an order of magnitude compared to our previous studies 
\cite{Coldea97}. The detectors form a semicircular 51-element array covering
a wide range of scattering angles ($25.75^{\circ }$ to $158.0^{\circ })$.
The sample was mounted with the ($a,b$) scattering plane horizontal in a
dilution refrigerator insert with base temperature 0.1 K.

The results for the ${\bf b}^{\ast }$ dispersion are shown in Fig.\ \ref
{fig2}(a). Although the data points also have a finite wavevector component
along ${\bf a}^{\ast }$, no measurable dispersion could be detected along
this direction confirming that the coupling between layers $J^{\prime \prime
}$ is negligible. The observed dispersion is well accounted for by the
principal spin-wave mode \cite{Jensen91} of Hamiltonian (\ref{hamiltonian}), 
$\omega ({\bf k})=\sqrt{\left( J_{{\bf k}}-J_{{\bf Q}}\right) \left[ \left(
J_{{\bf k}-{\bf Q}}+J_{{\bf k}+{\bf Q}}\right) /2-J_{{\bf Q}}\right] }$
where the Fourier transform of exchange couplings is $J_{{\bf k}}=\tilde{J}%
\cos {2\pi k}+2\tilde{J^{\prime }}\cos {\pi k} \cos {\pi l}$ and ${\bf k} =
(h , k ,l)$. The ordering wavevector in the 2D Brillouin zone of the
triangular lattice is ${\bf Q}$=$(0.5+\epsilon _{0}) {\bf b^{\ast}}$ and the
effective exchange parameters $\tilde{J}$=0.62(1) meV and $\tilde{J}^{\prime
}=2\tilde{J}\sin \pi \epsilon _{0}$ (fixed) are in agreement with \cite
{Coldea97}. The quantum renormalization of the excitation energy $\tilde{J}
/J$=1.65 is very large and is similar to the exact result $\pi /2$ for the
1D $S$=1/2 HAFC (see e.g. \cite{Muller81}). In contrast, the spin-wave
velocity (energy)\ renormalization in the unfrustrated $S$=1/2 HSL is only
1.18. {\em Such large renormalizations of energy (1.65) and incommensuration
(0.56) show the crucial importance of quantum fluctuations in the low-field
state of Cs$_{2}$CuCl$_{4}$}.

A remarkable feature of the measured dynamical correlations is that these do
not show single particle poles, but rather extended continua. Fig.\ \ref
{fig2}(b) (open circles) shows a scan at the magnetic zone boundary taken at
0.1 K. The scattering is highly asymmetric with a significant high-energy
tail. The non-magnetic background (dashed line) is modeled by a
constant-plus-exponential function. The magnetic peak disappears at 15 K
(solid circles) and is replaced by a broad, overdamped, paramagnetic signal.
Figs.\ \ref{fig3}A-D show 0.1 K data properly normalized and corrected for
absorption, and with the non-magnetic background subtracted. To quantify
discussion of the dynamical correlations measured by neutron scattering, we
first consider a spin-wave model, which is known to provide a good
description of the unfrustrated HSL \cite{Hayden91}.

The dynamical correlations of the spin-wave model \cite{Jensen91} for
Hamiltonian (\ref{hamiltonian}) exhibit single-particle poles from three
spin-1 magnon modes, polarized with respect to the cycloidal plane. Fig.\ 
\ref{fig2}(a) shows the main dispersion mode $\omega (k)$, polarized
out-of-plane, and the two secondary modes, $\omega ^{-}(k)=\omega ( k - Q)$
and $\omega ^{+}( k)=\omega (k + Q)$, both polarized in-plane, where the
equilibrium spin direction rotates in-plane with wavevector $Q$. The
expected scattering is given by the dashed lines in Figs.\ \ref{fig3}A-D,
which clearly fails to account for the observed intensity as well as the
extended high-energy tail of the scattering. This tail is not an
instrumental effect as the FWHM of the energy resolution (horizontal bar in
Fig.\ \ref{fig2}(b)) is an order of magnitude narrower than the signal
width. Including next-order processes also fails to account for the
scattering: the two-magnon scattering (polarized in-plane) contribution to
the lineshape is also shown in Fig.\ \ref{fig3}D (shaded area) - this was
calculated numerically using the method described in \cite{Tennant95b} using
the experimentally estimated spin reduction $\Delta S\sim $ 0.13 \cite
{Coldea96} to normalize the elastic, one- and two-magnon scattering.

Because in a neutron scattering process the total spin changes by $\Delta
S_{total}$=0,$\pm 1$ the absence of single-particle poles and the presence
of excitation continua implies that the underlying excitations carry
fractional quantum numbers. For $J^{\prime }$=0, these are rigorously known
to be $S$=1/2 spinons \cite{Muller81}, and two-spinon production is the
principal neutron scattering process \cite{Tennant95a}. Our analysis shows
that the measured scattering can be described by the M\"{u}ller ansatz
lineshape appropriate to the 1D $J^{\prime }=0$ limit $S(k,\omega )\sim
\Theta \left( \omega -\omega _{l}(k)\right) \Theta \left( \omega
_{u}(k)-\omega \right) /\sqrt{\omega ^{2}-\omega _{l}^{2}(k)}$, where $%
\Theta $ is the Heaviside step function, and $\omega _{l}$ and $\omega _{u}$
are the lower and upper continuum boundaries, generalized to 2D such that:
(1) the total cross-section has three continua with the lower boundaries $%
\omega (k)$, $\omega ^{-}(k)$ and $\omega ^{+}(k)$ shown in Fig.\ \ref{fig2}%
(a); (2) the continua have equal weights and are isotropic in spin space;
and (3) a modified upper boundary $\omega _{u}$ (dashed upper line in Fig.\ 
\ref{fig2}(a)) is used. This model provides an excellent description of the
data, see Figs.\ \ref{fig3}A-D. It is also noteworthy that both the
asymmetric dispersion, characteristic of 2D frustrated couplings, and the
excitation continua are essentially unchanged at $T$=0.9 K above $T_{N}$%
=0.62 K in the disordered spin liquid phase showing that ordering affects
only the low-energy behavior. We conclude that, {\em in contrast to the HSL (%
$J$=0) where unfrustrated couplings confine spinons into $S$=1 magnons
throughout the Brillouin zone, Cs$_{2}$CuCl$_{4}$ has fractional spin
quasiparticles carrying the same quantum numbers as in the HAFC ($J^{\prime
} $=0), namely $S$=1/2 spinons, and, further, that these spinons are
modified by the two-dimensionality at all energy scales}. Although no
evidence of spinon confinement is observed at any of the energy scales
probed in our experiments, low energy $S$=1 Goldstone modes are expected to
occur in the 3D ordered phase; weakly coupled HAFCs have recently been shown
to exhibit dimensional crossover in the dynamical correlations from
low-energy 3D spinwaves to high-energy 1D spinon continua on an energy scale
of the interchain coupling \cite{Essler97}.

Magnetic fields applied within the ordering plane have a profoundly
different effect from those along $a$. In fields along $c$ a transition 
occurs above 1.4 T ($T<$0.3 K) to a phase, marked S on Fig. 1(b), where 
(1) the structure is elliptical with a large elongation along the field 
direction, (2) the incommensuration approaches a linear relation with 
field with a large slope \cite{longpaper}, see Fig. 1(e), and (3) the 
total ordered moment decreases with increasing field. Above 2.1 T there 
is no long-range order at least down to 35 mK, and the system is in a 
spin liquid state. In this phase the dynamical correlations show shifts 
of continua and redistribution of scattering weight \cite{Coldea97} as 
expected for spinon states and compared in \cite{Coldea97} with 1D results 
\cite{Muller81}. However, 2D correlations are also important (by continuity 
they persist to the ferromagnetic F phase at saturation and to the spin 
liquid phase above $T_{N}$, both show 2D character) and give rise to 
an asymmetric distribution of scattering weight around $k$=1.5 
\cite{Coldea97,longpaper}. Linear field dependence of the incommensuration 
of the {\em longitudinal} spin correlations is a signature of exclusion 
statistics for the spinon quasiparticles in the HAFC, and the observed 
similar behavior of the incommensuration in the S phase dominated by the 
ordering of the {\em longitudinal} spin components suggests that 
{\em exclusion statistics are important for the quasiparticles in 
Cs$_{2}$CuCl$_{4}$}. The existence of a modulated continuum upper boundary 
indicating phase space restrictions for paired states also supports this 
conclusion. Susceptibility measurements \cite{longpaper} show no evidence 
of a phase transition between the spin liquid behavior in zero field above 
$T_{N}$=0.62 K and the disordered phase found for fields greater than 2.1 T 
down to at least 35 mK.{\em This indicates that fields applied within 
the ordering plane stabilize the fractional spin liquid state}.

In conclusion we have studied the ground-state and excitations of the
frustrated quantum antiferromagnet Cs$_{2}$CuCl$_{4}$. This material has a
2D Hamiltonian interpolating between the square, triangular and 1D
Heisenberg antiferromagnets, and shows: (1) very strong quantum
renormalizations indicating the importance of fluctuations; (2) continua in
the dynamical correlations demonstrating fractional excitations; (3) very
large field-driven incommensuration and disorder effects from inplane fields
showing that exclusion statistics are important; and (4) stabilization of a
spin liquid ground state by inplane fields. We believe new theoretical work
is needed to explain these findings.

Full details of the analysis and extensive results from other related
experiments on Cs$_{2}$CuCl$_{4}$ will be given in a forthcoming publication 
\cite{longpaper}. We would like to thank M. Eskildsen, M.A. Adams and M.J.
Bull for technical support and we acknowledge very useful discussions with
D.F. McMorrow, R.A. Cowley, F.H.L. Essler, A.O. Gogolin and M. Kenzelmann.
ORNL is managed for the US DOE by UT-Battelle, LLC, under contract
DE-AC05-00OR22725. A.M.T. is grateful to Isaac Newton Institute and
Trinity College Cambridge for kind hospitality.

\begin{figure}[t]
\caption{(a) Dispersion of the magnetic excitations along the $b^*$-axis ($T$%
=0.1 K, zero field). Filled symbols (triangles from \protect\cite{Coldea97})
show the main peak in the measured lineshape and solid line is a fit to the
principal spin-wave mode $\protect\omega(k)$ of Hamiltonian (\ref
{hamiltonian}) (dashed and dash-dotted lines show the corresponding
dispersion of the other two secondary modes, $\protect\omega^+$ and $\protect%
\omega^-$ respectively, as described in the text). Typical scan trajectories
are shown by the light shaded regions labelled {\bf A}-{\bf D} (the line
thickness is the wavevector averaging) and the measured data are shown in
Fig.\ \ref{fig3}{\bf A}-{\bf D}. Open circles show the experimentally
estimated upper boundary of the continuum and upper (heavy) dashed line is a
guide to the eye. Dotted area indicates extent of magnetic scattering. (b)
Intensity measured along scan {\bf D} above, in the cycloidal phase at 0.1 K
(open circles) and in the paramagnetic phase at 15 K (solid circles). Data
points are the raw counts and dashed line shows the estimated non-magnetic
background. Solid lines are guides to the eye and the horizontal bar
indicates the instrumental energy resolution.}
\label{fig2}
\end{figure}

\begin{figure}[b]
\caption{ Magnetic inelastic scattering measured at 0.1 K along the light
shaded regions labelled {\bf A}-{\bf D} in Fig.\ \ref{fig2}(a). Top axis
shows wavevector change along scan direction. Counting times were typically
35 hours at an average proton current of 170$\protect\mu$A. Solid lines are
fits to a modified two-spinon cross-section (see text). Vertical arrows
indicate estimated upper boundary. Dashed lines show predicted lineshape for
polarised cycloidal spin waves and the dark shaded region (shown only in 
{\bf D} for brevity) indicates the estimated two-magnon scattering
continuum. All calculations include the isotropic magnetic form factor of Cu$%
^{++}$ ions and the convolution with the spectrometer resolution function.}
\label{fig3}
\end{figure}


\begin{references}
\bibitem{Muller81}  G. M\"{u}ller {\em et al.}, 
Phys. Rev. B {\bf 24}, 1429 (1981); J.C. Talstra and F.D.M. Haldane, Phys.
Rev. B {\bf 50}, 6889 (1994).

\bibitem{Tennant95a}  D.A. Tennant {\em et al.}, 
Phys. Rev. B {\bf 52}, 13368 (1995); 
D.C. Dender {\em et al.}, 
Phys. Rev. B {\bf 53}, 2583 (1996). 

\bibitem{Anderson73}  P.W. Anderson, Mat. Res. Bull. {\bf 8}, 153 (1973).

\bibitem{Kalmeyer87}  V. Kalmeyer and R.B. Laughlin, Phys. Rev. Lett. {\bf 59%
}, 2095 (1987); 
Phys. Rev. B {\bf 39}, 11879 (1989);

\bibitem{larkin}  L. B. Ioffe and A. I. Larkin, Phys. Rev. B{\bf 39}, 8988
(1989).

\bibitem{lee}  P. A. Lee and N. Nagaosa, Phys. Rev. B{\bf 46}, 5621 (1992).

\bibitem{Hayden91}  S.M. Hayden {\em et al.}, Phys. Rev. Lett. {\bf 67},
3622 (1991); 
S.J. Clarke {\em et al.}, 
Solid State Commun. {\bf 112}, 561 (1999). 

\bibitem{Sachdev99}  See for example S. Sachdev, in {\em Quantum Phase
Transitions} (Cambridge, Cambridge,1999) ch.13.

\bibitem{Watanabe99}  S. Watanabe and H. Yokoyama, J. Phys. Soc. Japan {\bf %
68}, 2073 (1999); 
D. Allen, F.H.L. Essler, and A.A. Nersesyan, Phys. Rev. B {\bf 61}, 8871
(2000). 

\bibitem{Coldea97}  R. Coldea {\em et al.}, 
Phys. Rev. Lett. {\bf 79}, 151 (1997). 

\bibitem{Coldea96}  R. Coldea {\em et al.}, 
J. Phys.:Condens. Matter {\bf 8}, 7473 (1996); 
J. Magn. Magn. Mater. {\bf 177}, 659 (1998). 

\bibitem{Wen89}  X.-G. Wen, F. Wilczek, and A. Zee, Phys. Rev. B {\bf 39},
11413 (1989); 
R.B. Laughlin, Science {\bf 242}, 525 (1988). 

\bibitem{longpaper}  R. Coldea {\em et al.}, 
(in preparation).

\bibitem{Weihong99}  Z. Weihong, R.H. McKenzie, and R.R.P. Singh, Phys. Rev.
B {\bf 59}, 14367 (1999). 

\bibitem{Jensen91}  J. Jensen and A.R. Mackintosh, in {\em Rare Earth
Magnetism. Structures and Excitations} (Clarendon, Oxford, 1991) p. 291.

\bibitem{Tennant95b}  D.A. Tennant {\em et al.}, 
Phys. Rev. B {\bf 52}, 13381 (1995). 

\bibitem{Essler97}  H.J. Schulz, Phys. Rev. Lett. {\bf 77}, 2790 (1996); 
F.H.L. Essler, A.M. Tsvelik, and G. Delfino, Phys. Rev. B{\bf 56}, 11001
(1997); 
B. Lake, D.A. Tennant, and S.E. Nagler, Phys. Rev. Lett. {\bf 85}, 832 (2000). 

\begin{figure}[b]
\caption{(a) 2D couplings in Cs$_{2}$CuCl$_{4}$: Strong bonds $J$ (heavy
lines) and smaller frustrating zig-zag bonds $J^{\prime }$ (thin lines). (b)
and (c) Magnetic phase diagram in a field along the $c$ and $a$ axes,
respectively. Symbols show the boundaries of the various phases described in
the text measured using neutron scattering (squares) \protect\cite
{Coldea96,longpaper} and susceptibility (circles) \protect\cite{longpaper}.
Solid curves are a guide to the eye, and dashed line indicates crossover to
paramagnetic behavior. (d) Spin rotation in the ($b,c$) plane. Heavy arrows
are spin vectors and the circle indicates the spin rotation upon translation
along the $b$-axis. (e) Incommensuration $\protect\epsilon$ vs. field along 
$c$ \protect\cite{longpaper}. (f) $\protect\epsilon$ vs. field along $a$ 
(solid line is a guide to the eye, solid circle is from \protect\cite{longpaper}).}
\label{fig1}
\end{figure}
\end{references}
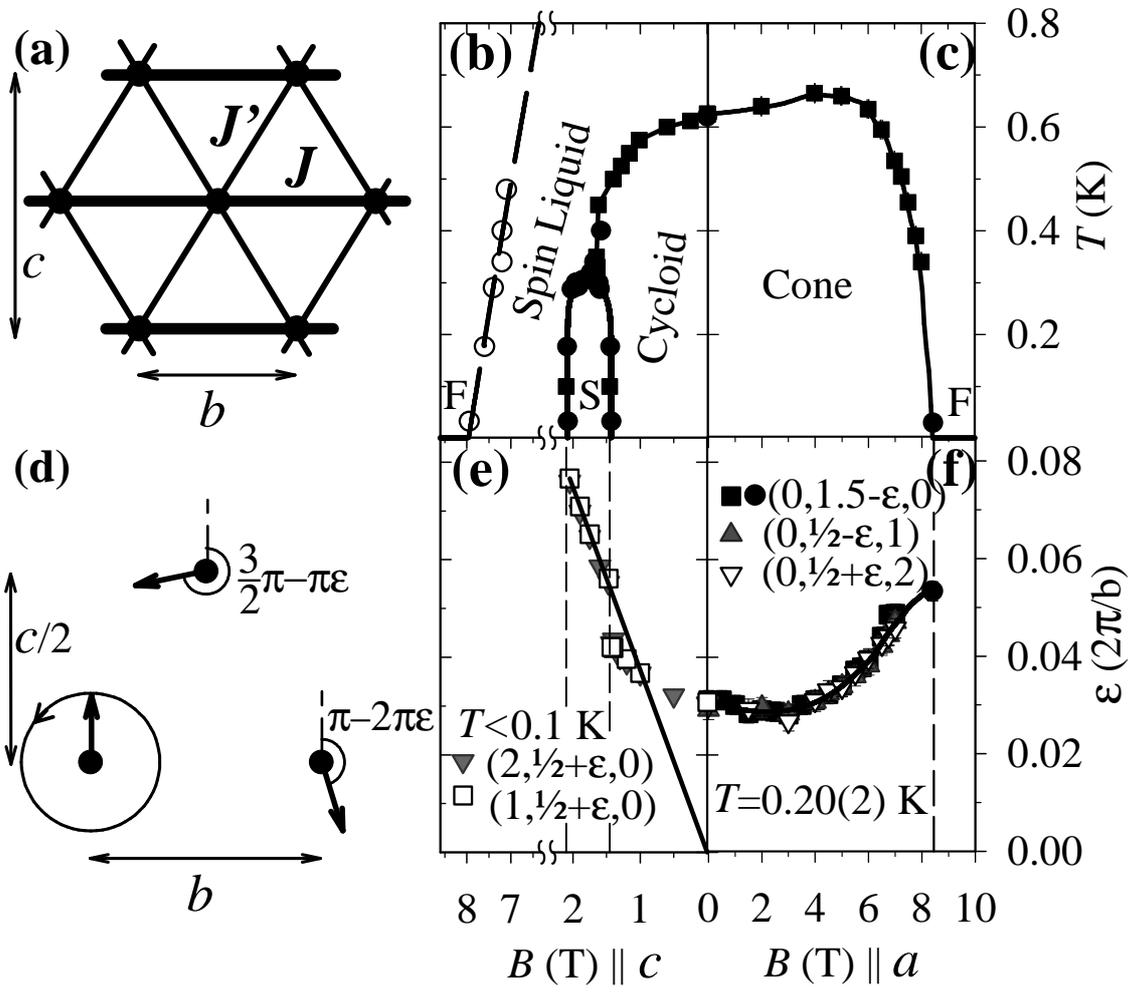
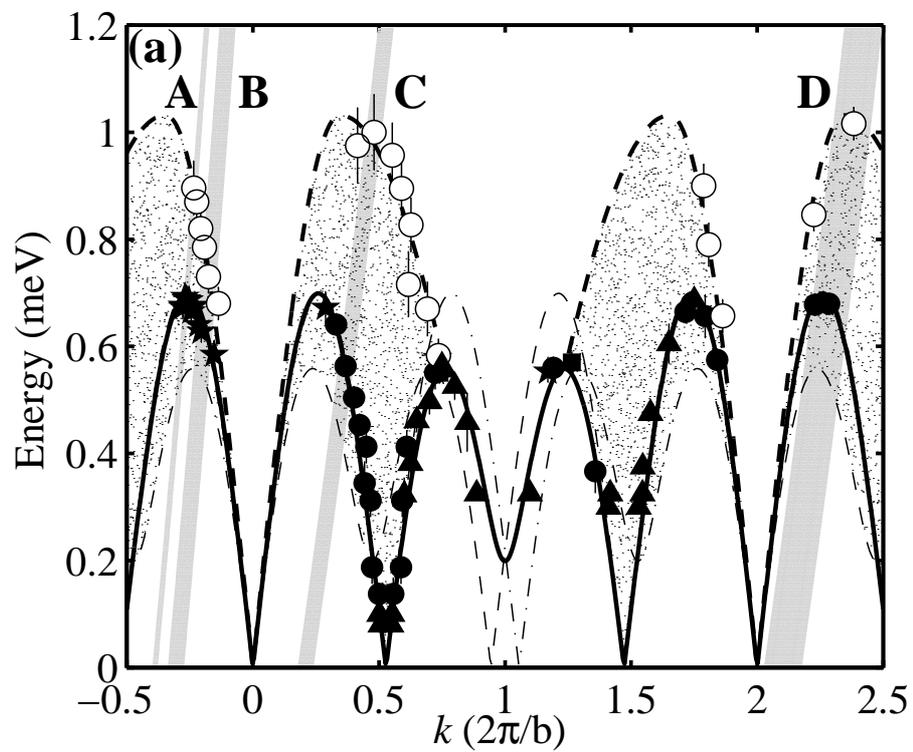
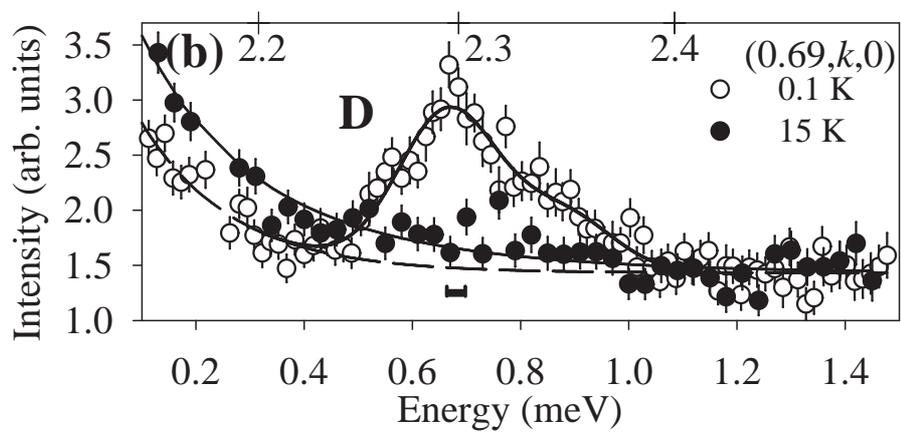
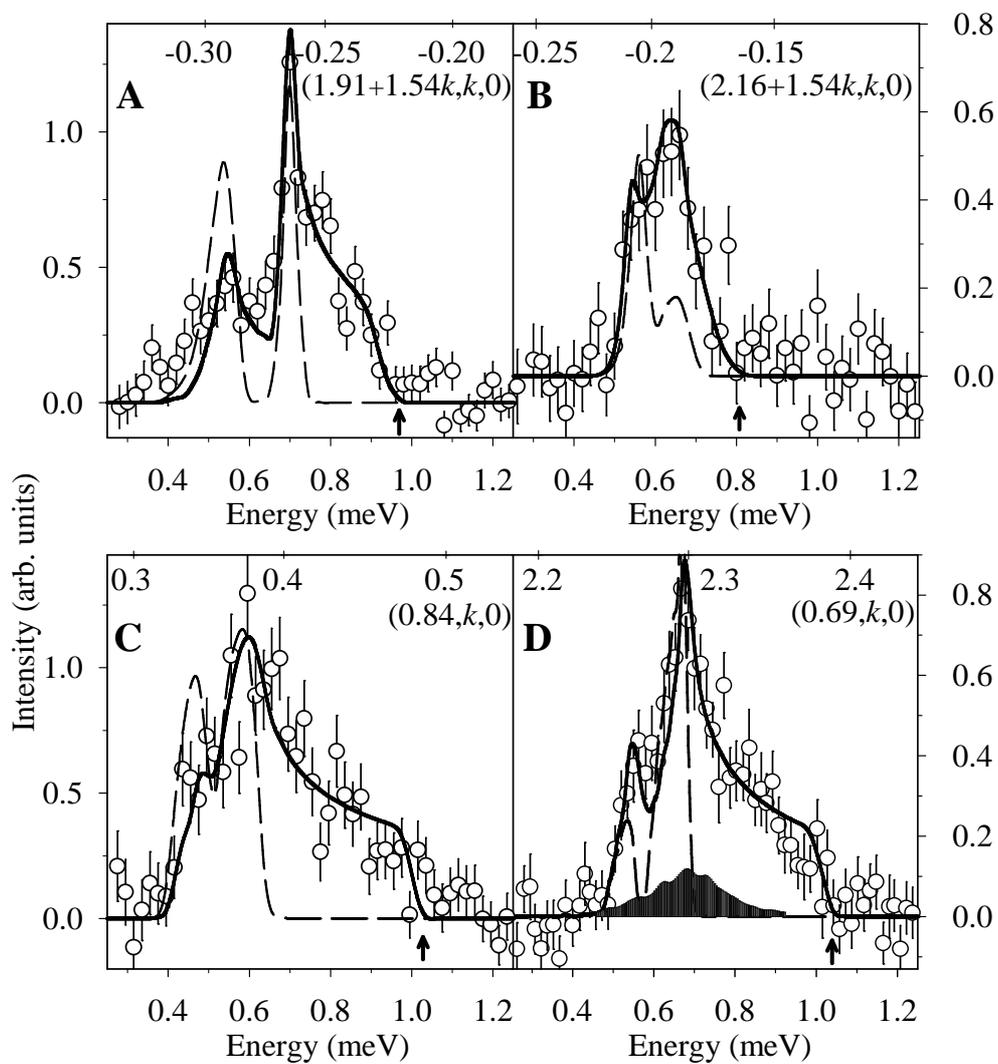
\end{document}